\documentclass[12pt]{article}
\pdfoutput=1

\input epsf
\usepackage{amssymb,amsmath,mathtools,wrapfig,subfigure}
\usepackage[matrix,arrow,curve]{xy}
\usepackage{cite}
\usepackage{graphicx,color}
\usepackage[debug,pageanchor=false]{hyperref}
\definecolor{link}{rgb}{.8,.15,.1}
\hypersetup{colorlinks=true,linkcolor=link,citecolor=link,urlcolor=link,linktocpage}

\makeatletter
\@addtoreset{equation}{section}
\makeatother

\def\rr {{\Bbb R}}

\begin{document}

\begin{titlepage}

\begin{center}

\vskip .3in \noindent



{\Large \textbf{Generalized geometry of two-dimensional vacua}}
\bigskip

Dario Rosa \\

\bigskip
  Dipartimento di Fisica, Universit\`a di Milano--Bicocca, I-20126 Milano, Italy\\
and\\
INFN, sezione di Milano--Bicocca,
I-20126 Milano, Italy\\
dario.rosa@mib.infn.it


\vskip .5in
{\bf Abstract }

\end{center}

We derive the conditions for unbroken supersymmetry for a Mink$_2$, $(2,0)$ vacuum, arising from Type II supergravity on a compact eight-dimensional manifold $\mathcal{M}_8$. When specialized to internal manifolds enjoying $SU(4) \times SU(4)$ structure the resulting system is elegantly rewritten in terms of generalized complex geometry. This particular class of vacua violates the correspondence between supersymmetry conditions and calibrations conditions of D branes (supersymmetry-calibrations correspondence). Our analysis includes and extends previous results about the failure of the supersymmetry-calibrations correspondence, and confirms the existence of a precise relation between such a failure and a subset of the supersymmetry conditions.  

\vskip .1in


\noindent

\vfill
\eject

\end{titlepage}

\hypersetup{pageanchor=true}


\section{Introduction} 
\label{sec:intro}

The search for supersymmetric vacuum configurations in Type II supergravity has represented in the past a fruitful area of interplay between mathematics and physics. On the one hand the study of string theory compactifications to four dimensions in the absence of fluxes has driven to a lot of efforts in studying Calabi-Yau manifolds, leading to important progresses in understanding the geometrical properties of these spaces; on the other hand generalized complex geometry \cite{Hitchin:2004ut}, \cite{Gualtieri:2003dx} proved to be a powerful tool in order to study the more complicated (and interesting) story of vacuum solutions in presence of fluxes.

Generalized complex geometry was applied to the study of vacua in presence of fluxes for the first time in \cite{gmpt2}. In that paper the authors found that the conditions for unbroken supersymmetry for four-dimensional $\mathcal{N} = 1$ vacua are elegantly rewritten in terms of first-order differential equations involving a pair of pure spinors on the generalized internal tangent bundle $T_6 \oplus T_6^\ast$; this fancy formulation allowed to find a large number of explicit vacuum solutions.

A very interesting (and perhaps unexpected) observation was done in \cite{Martucci:2005ht}: it was found that the conditions for a Mink$_4$ vacuum, when expressed in the pure spinors formulation, are in one-to-one correspondence with the differential conditions satisfied by the calibration forms for {\it all} the admissible, static, magnetic D-branes in such a background.\footnote{An analogous story holds also for AdS$_4$ vacua \cite{koerber-martucci-ads}.} It is natural to ask whether the correspondence (which we will call the supersymmetry-calibrations correspondence) is also valid in more general situations and can be applied in dimensions different than four.

Motivated by this question, in \cite{Lust:2010by} it was checked that the supersymmetry-calibrations correspondence continues to hold also for Mink$_6$ vacua preserving eight real supercharges, and this led to formulate the following conjecture: the supersymmetry-calibrations correspondence is valid for all Mink$_d$ vacua (with $d$ even) preserving a Weyl spinor on the external manifold. 

Specializing the discussion to the case of Mink$_2$, $\mathcal{N}=(2,0)$ vacua, the authors of \cite{Lust:2010by} conjectured that the conditions for unbroken supersymmetry should be
\begin{align}
\label{eq:exteriorpureintro}
  & d_{H} (e^{2A-\phi} \mathrm{Re} \psi_1) = \pm \frac {\alpha}{16} e^{2A} \ast_8 \lambda (f) \ , \nonumber \\
  & d_{H} (e^{2A -\phi} \psi_2) = 0 \ ,
\end{align}
where $d_H \equiv d - H \wedge$ and $\psi_1 = \frac{1}{e^A} \eta^1_+ \eta^{2 \dagger}_\mp$, $\psi_2 = \frac{1}{e^A} \eta^1_+ \eta^{2 c \dagger}_\mp$ are polyforms constructed as bilinears of the internal SUSY parameters $\eta^1_+$ and $\eta^2_\pm$ (which are Weyl spinors); finally the upper (lower) sign is for IIA (IIB). It is worth emphasizing that the correspondence was formulated for $\eta^1_+$ and $\eta^2_\pm$ being {\it pure} spinors on the internal manifold;\footnote{Contrary to what happens in lower dimensions, in eight dimensions not every Weyl (not Majorana) spinor is pure: as reviewed in section \ref{sec:generalizedhodgepure} an eight-dimensional Weyl spinor is pure if and only it satisfies an additional algebraic condition (\ref{eq:puritycondition}). From this it follows that the situation considered in \cite{Lust:2010by} is not the most general one for a Mink$_2$, $\mathcal{N} = (2,0)$ vacuum.} this assumption implies that the structure group on the generalized tangent bundle $T_8 \oplus T^\ast_8$ reduces to $SU(4) \times SU(4)$. 

In \cite{Prins:2013koa} it has been shown, by  making the additional assumption  that $\eta^1_+$ and $\eta^2_+$ are {\it proportional}, that in type IIB the conjecture of \cite{Lust:2010by} fails to be valid: the authors indeed have shown that the equations (\ref{eq:exteriorpureintro}) are not completely equivalent to supersymmetry and that they must be completed, in this particular case, with the condition 
\begin{equation}
\label{eq:pairingpurerewrittenintro}
   d_{H}^{\mathcal{J}_2} (e^{-\phi} \mathrm{Im} \psi_1) = - \frac{\alpha}{16} f \ ,
\end{equation}
where $d_H^{\mathcal{J}_2} \equiv [d_H , \mathcal{J}_2 \cdot ]$ (used for the first time in physical context in \cite{Tomasiello:2007zq}), and $\mathcal{J}_2$ is the generalized almost complex structure associated to the pure spinor on the generalized tangent bundle $\psi_2$ (further details are given in section \ref{sec:generalizedhodgepure}). They also gave a geometrical interpretation of this equation in terms of calibrations, motivated by the results obtained in dimensions greater than $2$.

It should be noted that the assumption that $\eta^1_+$ and $\eta^2_\mp$ are proportional is a strong assumption and can be made only in type IIB, since in type IIA $\eta^1_+$ and $\eta^2_-$ have different chiralities and cannot be proportional. In this case we have a reduction of the structure group to {\it strict} $SU(4)$ and this allows the use of ordinary complex geometry instead of generalized complex geometry. 

The authors of \cite{Prins:2013koa} conjectured that the final result does not change by removing the assumption of proportionality between $\eta^1_+$ and $\eta^2_+$, but they did not test this final statement; however they suggested that the ten-dimensional system found in \cite{Tomasiello:2011eb} could be useful in order to show such a conjecture. In this paper, motivated by the elegant result obtained in\cite{Prins:2013koa} and using the strategy that the authors conjectured to be useful, we remove the assumption of proportionality between $\eta^1_+$ and $\eta^2_\mp$ and show the validity of the results of \cite{Prins:2013koa} in the general case with non proportional spinors. 

As a further generalization we will show that the conditions for unbroken supersymmetry can be recast in an elegant form every time $\mathcal{M}_8$ enjoys an $SU(4) \times SU(4)$ structure, no matter whether $\eta^1_+$ and $\eta^2_\mp$ are pure or not. In other words, we will see that the conditions for unbroken supersymmetry take an elegant formulation if we assume that it exists a pair of pure spinors $\tilde{\eta}^1_+$ and $\tilde{\eta}^2_\mp$ (which in general will not coincide with the SUSY parameters $\eta^1_+$ and $\eta^2_\mp$).

Here is a description of the strategy that we will follow. We begin by considering the ten-dimensional system given in \cite{Tomasiello:2011eb}. This system gives a set of differential equations, written in terms of differential forms using again generalized complex geometry on $T_{10} \oplus T^\ast_{10}$, which an {\it arbitrary} ten-dimensional configuration must satisfy in order to preserve supersymmetry. Some of the equations (the so-called symmetry equations (\ref{eq:LK}) and the exterior equation (\ref{eq:psp10})) are concise and reminiscent of the pure spinor equations for four-dimensional vacua. Unfortunately they are not in general sufficient for supersymmetry to hold, and must be completed with the so-called pairing equations (\ref{eq:++1}) and (\ref{eq:++2}); these last equations are much more cumbersome than the others and they involve additional geometrical quantities that are not defined by the SUSY parameters $\epsilon_1$ and $\epsilon_2$. In some situations they are redundant (for example for four-dimensional and six-dimensional Minkowski vacua), but in general they are non trivial.

Having at disposal the ten-dimensional system (here reviewed in section \ref{sec:SUSY10d}), the only thing that we need to do is to specialize it to the case of a Mink$_2$, $\mathcal{N} = (2,0)$ vacuum. In this way we easily obtain (in section \ref{sec:factorization}) the conditions for supersymmetry for a {\it general} Mink$_2$, $\mathcal{N}= (2,0)$ vacuum, without making any assumption about the internal SUSY parameters $\eta^1_+$ and $\eta^2_\mp$. In other words, the equations that we obtain in this section are more general than the ones in \cite{Lust:2010by} (and refined in \cite{Prins:2013koa}) and can be applied also for vacua in which the internal SUSY parameters are not pure spinors on the internal manifold. 

As a further step we impose the condition that both $\eta^1_+$ and $\eta^2_\mp$ are pure (but not necessarily proportional) and, using generalized complex geometry and more specifically the so-called generalized Hodge diamond, we deduce that the exterior equation (\ref{eq:psp10}) simply reproduces both the equations (\ref{eq:exteriorpureintro}) conjectured in \cite{Lust:2010by} on the basis of the supersymmetry-calibrations correspondence, whereas the pairing equations are completely equivalent to (\ref{eq:pairingpurerewrittenintro}). This concludes the proof of the validity of the system found in \cite{Prins:2013koa} also in the case of $SU(4) \times SU(4)$ structure. It is worth noting that, using this strategy, the computations to obtain the result are much simpler than the ones used in \cite{Prins:2013koa} for the case of strict $SU(4)$ structure, and this represents one of the main advantages of the system \cite{Tomasiello:2011eb}. We note also that, in the limit of vanishing fluxes, equations (\ref{eq:exteriorpureintro}) and (\ref{eq:pairingpurerewrittenintro}) give the well-known results of Calabi-Yau four-fold for Type IIB and manifolds with $G_2$ holonomy for Type IIA.

Summarizing this first part of the work, we show that the exterior equation simply reproduces both the equations conjectured in \cite{Lust:2010by} using calibrations, whereas the pairing equations reproduce (\ref{eq:pairingpurerewrittenintro}), which expresses the failure of the correspondence. By combining this observation with the fact that in $d=4$ and in $d=6$ the correspondence holds and the pairing equations are completely redundant, it becomes reasonable to conjecture that the pairing equations parametrize the failure of the correspondence. Indeed the validity of this last statement is shown in full generality in \cite{Martucci:2011dn}: in that work it is shown that the symmetry equations (\ref{eq:LK}) and the exterior equation (\ref{eq:psp10}) are the {\it only} equations of the ten-dimensional system necessary to identify the calibrations of D branes and F1 strings; on the other hand the pairing equations have nothing to do with calibrations (of D branes and F1 strings at least).

Finally, in the last part of the paper, we remove the condition that $\eta^1_+$ and $\eta^{2}_\mp$ are pure spinors but we continue to assume that a pair of pure spinors $\tilde{\eta}^1_+$ and $\tilde{\eta}^2_\mp$ exists. Such an assumption allows us to rewrite the equations for unbroken supersymmetry in terms of the pure spinors on the generalized tangent bundle $\tilde{\psi}_1$ and $\tilde{\psi}_2$. We show that the exterior equation (\ref{eq:psp10}), when rewritten in terms of $\tilde{\psi}_1$ and $\tilde{\psi}_2$, continue to reproduce the equations (\ref{eq:exteriorpureintro}) without any modifications. The pairing equations (\ref{eq:++1}), (\ref{eq:++2}) are different from (\ref{eq:pairingpurerewrittenintro}) but nevertheless they continue to have an elegant formulation (equation (\ref{eq:pairingimpuremainfinal})), similar to (\ref{eq:pairingpurerewrittenintro}) but deformed with additional pieces. Therefore our analysis reveals that the exterior equation in the system (\ref{eq:susy10}) is sensible {\it only} to the condition of $SU(4) \times SU(4)$ structure, whereas the pairing equations takes into account whether $\eta^1_+$ and $\eta^2_\mp$ are pure or not. 

The paper is organized as follow. In section \ref{sec:2dgeometry} we discuss our spinorial Ansatz and the geometry defined by a Weyl spinor in two dimensions. In section {\ref{sec:SUSYgeneral} we discuss the ten-dimensional system \cite{Tomasiello:2011eb} and we specialize it to the case of Mink$_2$, $\mathcal{N} = (2,0)$ vacua. In section \ref{sec:pure} we specialize further the system by requiring that the internal SUSY parameters are pure spinors on the internal manifold. In section {\ref{sec:impure} we remove the assumption that the internal SUSY parameters are pure spinors and in section \ref{sec:conclusions} some conclusions and future projects are outlined. Finally, in the appendices we give some  technical details about the computations of the main text.

\section{Spinorial Ansatz and two-dimensional geometry}
\label{sec:2dgeometry}

In this section we will discuss how the ten-dimensional SUSY parameters $\epsilon_1$ and $\epsilon_2$ decompose in order to have an $\mathcal{N}=(2,0)$, $\mathrm{Mink}_2$ vacuum, namely a configuration of the form $\mathrm{Mink}_2 \times \mathcal{M}_8$ (with $\mathcal{M}_8$ compact) enjoying the maximal symmetry of $\mathrm{Mink}_2$ and where two real supercharges are preserved. We will also describe what kind of geometrical quantities are defined by a single Weyl (Not Majorana) spinor $\zeta$ in two dimensions.

\subsection{Spinorial Ansatz}
\label{sub:spinans}

We consider configurations with a metric of the form
\begin{equation}
\label{eq:metric}
  ds^2_{10} (x,y) = e^{2A (y)} ds^2_{\mathrm{Mink}_2} (x) + ds^2_{\mathcal{M}_8} (y) \ ,
\end{equation}
$x^\mu$ are the coordinates on $\mathrm{Mink}_2$ and $y^m$ are the coordinates on the internal manifold $\mathcal{M}_8$. As usual for vacuum solutions the manifold is given by a simple product $\mathcal{M}_{10} = \mathrm{Mink}_2 \times \mathcal{M}_8$, and the external part of the metric depends on the internal coordinates via the so-called warping factor $A(y)$ only. 

Moreover we are interested in $\mathcal{N} =(2,0)$ configurations, i.e. configurations like (\ref{eq:metric}) preserving supersymmetry for {\it any} given two-dimensional, complex, Weyl spinor $\zeta$. Therefore the ten-dimensional SUSY parameters $\epsilon_1$ and $\epsilon_2$ take the form
\begin{align}
\label{eq:spinans}
  & \epsilon_1 = \zeta \eta_+^1 + \zeta^c \eta_+^{c \,1}  \ ,\nonumber \\
  & \epsilon_2 = \zeta \eta_{\mp}^2 + \zeta^{c} \eta_{\mp}^{c\, 2} \ , 
\end{align}
where the upper sign is for IIA, the lower for IIB. $\zeta$ denotes a Weyl spinor (of positive chirality) in two dimensions and $\eta^i_{\pm}$ are two Weyl spinors on $\mathcal{M}_8$.\footnote{We will work with real gamma matrices both in $\mathrm{Mink}_2$ and in $\mathcal{M}_8$; such a basis in eight dimensions can be defined in terms of octonions \cite{Bryant}. Therefore the Majorana conjugates $\zeta^c$ and $\eta_\pm^{c\,i}$ are just the naive conjugates $(\zeta)^\ast$ and $(\eta^i_\pm)^\ast$.} Since we are not imposing also a Majorana condition on $\zeta$ (recall that in two dimensions Majorana-Weyl spinors can be defined) we see that $\zeta$ defines two real supercharges in two dimensions and (\ref{eq:metric}) is an $\mathcal{N} = (2,0)$ vacuum. Similarly to  (\ref{eq:spinans}), the ten-dimensional gamma matrices $\Gamma_M$ decompose as
\begin{equation}
\label{eq:gammadec}
   \Gamma_{\mu} = e^A \gamma_{\mu} \otimes 1 \ , \qquad \Gamma_m = \gamma^{(2)} \otimes \gamma_m \ ,
\end{equation}
where $\gamma_\mu$ and $\gamma_m$ are the real two-dimensional and eight-dimensional gamma matrices respectively, and $\gamma^{(2)}$ is the chiral operator in two dimensions. $M$ goes from $0$ to $9$.

To have a vacuum solution we need that the external spinor $\zeta$ satisfies a Killing spinor equation like
\begin{equation}
\label{eq:Killingmink}
  D_\mu \zeta = 0 \ .
\end{equation}
It is worth noting that a spinorial decomposition like (\ref{eq:spinans}) is not compatible with an $\mathrm{AdS}_2$ vacuum: indeed in this case the Killing spinor equation (\ref{eq:Killingmink}) becomes
\begin{equation}
\label{eq:KillingAdS}
  D_\mu \zeta_+ = \mu \gamma_\mu \zeta_- \ ,
\end{equation}
where $\zeta_+$ ($\zeta_-$) is a spinor of positive (negative) chirality, and $\mu$ is a constant proportional to the cosmological constant; it can be easily shown that $\zeta$ and $\zeta^c$ have the same chiralities and so we conclude that the spinorial Ansatz (\ref{eq:spinans}) is not compatible with (\ref{eq:KillingAdS}).

\subsection{Geometry defined by two-dimensional spinors}

Given the spinorial Ansatz (\ref{eq:spinans}) we want now to develop what kind of geometrical quantities can be defined using $\zeta$ and $\zeta^c$.

Given $\zeta$ of positive chirality we can introduce the barred spinor $\bar{\zeta} = \zeta^\dagger \gamma_0$ and a straightforward calculation shows that it has negative chirality. We can now define the bilinears $\zeta \otimes \bar \zeta$ and $\zeta \otimes \bar \zeta^c$ obtaining a couple of one-forms (or vectors), $z_\mu$ and $a_\mu$: 
\begin{align}
\label{eq:2dbilinears}
  & \zeta \otimes \bar \zeta = \frac 12 \bar{\zeta} \gamma_\mu \zeta \gamma^\mu = z_\mu dx^\mu \ , \nonumber \\
  & \zeta \otimes \bar \zeta^c = \frac 12 \bar{\zeta}^c \gamma_\mu \zeta \gamma^\mu = a_\mu dx^\mu \ ; 
\end{align}
our aim is now to understand the geometrical properties of both.

To start with, $z$ and $a$ are null: a simple Fierz computation gives\footnote{We will make systematically use of the Clifford map $dx^{m_1} \wedge \dots \wedge dx^{m_n} \rightarrow \gamma^{m_1 \dots m_n}$ to identify as usual forms with bispinors.}
\begin{equation}
\label{eq:2dnull}
  2z \zeta = \bar \zeta \gamma_\mu \zeta \gamma^\mu \zeta = \gamma^\mu \zeta \bar \zeta \gamma_\mu \zeta = 0 \ ,
\end{equation}
where we used the well-known relation $\gamma^\mu C_k \gamma_\mu = (-)^k(d - 2k) C_k$. From (\ref{eq:2dnull}) it follows $z^2=0$ and an identical computation shows that also $a$ is null. Moreover $z$ and $a$ are proportional since we have
\begin{equation}
\label{eq:za}
  z \zeta \bar \zeta^c = 0 \ ,
\end{equation} 
as an obvious consequence of (\ref{eq:2dnull}); recalling the formula $\gamma^\mu  C_k= (d x^\mu \wedge + g^{\mu\nu} \iota_\nu) C_k$, (\ref{eq:za}) can be rephrased as
\begin{equation}
 z \wedge a = z\llcorner a = 0 \ ,
\end{equation}
telling us that $a$ is proportional to $z$
\begin{equation}
\label{eq:zpropa}
  a = g(x) z \ .
\end{equation}
Finally, recalling that in two Lorentzian dimensions we have the identification
\begin{equation}
\label{eq:lambdadefinition}
  \gamma^{(2)} C_k = \ast_2 \lambda C_k \ , \qquad \lambda C_k \equiv (-1)^{\lfloor\frac 12 k \rfloor} C_k \ ,
\end{equation}
relating the action from the left of the chiral operator to the Hodge dual operator, we conclude that both $z$ and $a$ are self-duals
\begin{equation}
\label{eq:self}
 \ast_2 z = z \ , \qquad \ast_2 a = a \ .
\end{equation}

We can also determine the reality properties of these vectors. Evaluating the expression $\gamma^0 (\zeta \bar \zeta)^\dagger \gamma^0$ (and the analogous one including $\zeta \bar \zeta^c$) one deduces that $z$ is real and $a$ is complex.

To conclude this section we note that $z$ and $a$ are $d$-closed: indeed the external differential acts on a bispinor of odd degree as
\begin{equation}
\label{eq:derbispinor}
  d z = d (\zeta \bar{\zeta}) = \frac 12 \left[ \gamma^\mu , D_\mu \zeta \bar{\zeta} \right] \ , 
\end{equation}
and using (\ref{eq:Killingmink}) one obtains
\begin{equation}
\label{eq:zaclosed}
d z = d a = 0 \ .
\end{equation}

\section{Supersymmetry conditions: general discussion}
\label{sec:SUSYgeneral}
In this section we will review the conditions for unbroken supersymmetry in type II supergravities. We will then specialize them to two-dimensional $\mathcal{N} = (2,0)$ vacua obtaining a set of conditions for these particular backgrounds. Some of the equations will look a bit scary at first sight but in the next sections we will see that the situation is completely different when $\mathcal{M}_8$ enjoys an $SU(4) \times SU(4)$ structure.

\subsection{Review of the ten-dimensional system}
\label{sec:SUSY10d}
Let us review the conditions for unbroken supersymmetry in ten dimensions as derived in \cite{Tomasiello:2011eb}. All the material presented in this section is not new but we review it in order to have a self-contained discussion.

Using the ten-dimensional SUSY parameters $\epsilon_1$ and $\epsilon_2$ we can construct two different vectors (or equivalently one-forms)
\begin{equation}
\label{eq:K_idefinition}
  K^M_i \equiv \frac{1}{32} \bar{\epsilon}_i \Gamma^M \epsilon_i \ , \qquad K \equiv \frac 12 (K_1 + K_2) \ , \qquad \tilde K \equiv \frac 12 (K_1 - K_2) \ .
\end{equation}
We can also consider the polyform
\begin{equation}
\label{eq:Phidefinition}
  \Phi = \epsilon_1 \bar \epsilon_2  \ ,
\end{equation}
defining many different G-structures on the ten-dimensional tangent bundle, all of them corresponding to a single structure on the generalized ten-dimensional tangent bundle $T_{10} \oplus T^\ast_{10}$. The situation would appear to be completely analogous to what happens for four-dimensional $\mathcal{N}=1$ vacua, where the pure spinors $\phi_+$ and $\phi_-$ define together an $SU(3) \times SU(3)$ structure on the generalized tangent bundle of the internal manifold $T_6 \oplus T^\ast_6$, however one can show that $\Phi$ is {\it not} a pure spinor and as a consequence of this fact it is not sufficient to fully reconstruct the metric and the B-field. This feature forces us to introduce additional geometrical data and indeed in \cite{Tomasiello:2011eb} two additional vectors $e_{+1}$ and $e_{+ 2}$ satisfying 
\begin{equation}
 e_{+ i}^2 = 0 \ , \qquad e_{+i} \cdot K_i = \frac 12 \ , \qquad i=1,2 \ ,
\end{equation}
are introduced.

We can now reformulate the conditions for unbroken supersymmetry in terms of the geometrical data $(K, \tilde K, \Phi, e_{+i})$ just discussed, obtaining the following system 
\begin{subequations}\label{eq:susy10}
	\begin{align}
		\label{eq:LK}&L_K g = 0 \ ,\qquad d\tilde K = \iota_K H \ ;\\
		\label{eq:psp10}
		&d_H(e^{-\phi} \Phi) = -(\tilde K\wedge + \iota_K ) F \ ;\\
		& \label{eq:++1} 
		( e_{+1}\cdot \Phi \cdot e_{+2}\, ,\,\Gamma^{MN} [  \pm d_H(e^{-\phi} \Phi \cdot e_{+2}) + \frac 12 e^\phi d^\dagger(e^{-2\phi} e_{+2})\Phi - F ])=0\ ;\\
		& \label{eq:++2}
		( e_{+1}\cdot \Phi \cdot e_{+2}\, ,\, [   d_H(e^{-\phi} e_{+1} \cdot \Phi ) - \frac 12 e^\phi d^\dagger(e^{-2\phi} e_{+1})\Phi - F ] \Gamma^{MN})=0\ .
	\end{align}
\end{subequations}
((\ref{eq:LK}) already appeared in \cite{koerber-martucci-ads}, \cite{hackettjones-smith} and \cite{figueroaofarrill-hackettjones-moutsopoulos}.) Here, $\phi$ is the dilaton, $H$ is the NSNS three-form, and $d_H\equiv d-H\wedge$. $F$ is the total RR field strength $F=\sum_k F_k$ (where the sum is from $0$ to $10$ in IIA and from 1 to 9 in IIB), which is subject to the self-duality constraint 
\begin{equation}\label{eq:F*F}
	F=*_{10} \lambda (F) \ .
\end{equation}
$(\,,\,)$ is the ten-dimensional Chevalley-Mukai pairing of forms that, in $d$ dimensions, is defined by
\begin{equation}
   (\alpha, \beta) = (\alpha \wedge \lambda(\beta))_{d} \ , 
\end{equation} 
where $_d$ means that we keep only the d-form part, $\alpha$ and $\beta$ are two (poly)-forms and the $\lambda$ operator acts on a k-form $\alpha_k$ as defined in (\ref{eq:lambdadefinition}).

Equations (\ref{eq:susy10}) are \emph{necessary and sufficient} for supersymmetry to hold \cite{Tomasiello:2011eb}. To also solve the equations of motion, one needs to impose the Bianchi identities, which away from sources (branes and orientifolds) read 
\begin{equation}\label{eq:bianchi}
	dH=0 \ ,\qquad d_H F=0\ .
\end{equation}
It is then known (see \cite{lust-tsimpis} for IIA, \cite{gauntlett-martelli-sparks-waldram-ads5-IIB} for IIB) that almost all of the equations of motion for the metric and dilaton follow.

It should be noted that equations (\ref{eq:LK}) and (\ref{eq:psp10}) are very elegant: apart from the first equation in (\ref{eq:LK}) (expressing that $K$ has to be a Killing vector) they are formulated in terms of differential forms and exterior calculus only and they are much simpler to treat than the original SUSY conditions. Unfortunately, they are necessary to supersymmetry to hold but not sufficient and they must be completed with (\ref{eq:++1}) and (\ref{eq:++2}) (which we will call pairing equations). One can show that in some particular situations (\ref{eq:++1}) and (\ref{eq:++2}) can be dropped since they are redundant (this is the case for four-dimensional vacua, for example) but in general they carry additional content (examples of such situations can be found in \cite{Haack:2009jg}, \cite{Giusto:2013rxa} and \cite{Rosa:2013jja}). 

Of course it is possible that a better version of the pairing equations exists, and this is one of the aims of this paper: we will see that for $\mathcal{N} = (2,0)$, $\mathrm{Mink}_2$ vacua pairing equations can be rewritten in an elegant form if the internal manifold enjoys an $SU(4) \times SU(4)$ structure, in particular without making any use of the Chevalley-Mukai pairing and using only exterior calculus. It should be emphasized that this is not just a particular (and a particularly lucky) case: in \cite{Tomasiello:2011eb}, a decomposition of the SUSY parameters $\epsilon_i$ in terms of a two-dimensional Majorana-Weyl spinor $\zeta_i$ and of an eight-dimensional Majorana-Weyl spinor $\eta_i$ is often used. Therefore it is conceivable that some of the results presented in the following sections suggest a way to find a better formulation of the pairing equations in more general situations.

\subsection{Factorization}
\label{sec:factorization}
As explained in section \ref{sec:2dgeometry} we will consider backgrounds with a metric of the form (\ref{eq:metric}) and with a spinorial Ansatz like (\ref{eq:spinans}), we will also impose that our configuration is a vacuum, i.e. that the maximal symmetry of $\mathrm{Mink}_2$ is preserved by all the fields.

Given the spinorial Ansatz (\ref{eq:spinans}) we can immediately compute the polyform $\Phi$ (equation (\ref{eq:Phidefinition}))
\begin{align}
\label{eq:Phifactorized}
  \Phi & = \mp \bigl( (\zeta \bar \zeta)(\eta^1_+ \eta^{2 \dagger}_{\mp}) + (\zeta^c \bar \zeta) (\eta^{1 c} \eta^{2 \dagger}_{\mp}) + \mathrm{c.c.} \bigr) \nonumber \\
   & = \mp 2 \mathrm{Re} \left(e^A z \wedge \psi_1 + e^A a \wedge \psi_2 \right) \ , 
\end{align}
where the decomposition (\ref{eq:gammadec}) of the ten-dimensional gamma matrices is used. In (\ref{eq:Phifactorized}) $z$ and $a$ are the two-dimensional vectors defined in (\ref{eq:2dbilinears}), whereas with $\psi_1$ and $\psi_2$ we denote the eight-dimensional bilinears
\begin{equation}
\label{eq:psidefinition}
  \psi_1 \equiv \eta^1_+ \eta^{2 \dagger}_{\mp} \ , \qquad \psi_2 \equiv \eta^1_+ \eta^{2 c \dagger}_{\mp} \ .
\end{equation}
Notice that, since not every eight-dimensional Weyl spinor $\eta_+$ is pure, $\psi_1$ and $\psi_2$ are not in general pure spinors on the generalized eight-dimensional tangent bundle $T_8 \oplus T^\ast_8$ and so in general they do not induce a reduction of the structure group to $SU(4) \times SU(4)$. Further details about this point will be presented in section \ref{sec:pure}.

We need also the vectors $K$ and $\tilde K$ appearing in (\ref{eq:K_idefinition}). To this end we start by computing $K_1$ obtaining
\begin{align}
\label{eq:K_1}
  & 32 K_1 = e^{-A} \bigl[4 z ||\eta^1_+||^2 + 2 a (\eta^1_+)^2 + \mathrm{c.c} \bigr] \ , \qquad ||\eta^1_+||^2 \equiv \eta^{1 \dagger}_+ \eta^1_+ \ , \qquad (\eta^1_+)^2 \equiv (\eta^{1 t}_+ \eta^1_+) \, 
\end{align}
notice that $||\eta^1_+||^2$ is real, and $(\eta^1_+)^2$ is complex. If we now impose\footnote{When $\mathcal{M}_8$ is compact a famous no-go theorem requires the presence of sources with negative tension like orientifold planes \cite{deWit:1986xg}, \cite{Maldacena:2000mw}. The request that such orientifolds be supersymmetric imposes the conditions (\ref{eq:equalnorms}) that therefore has to be considered as a necessary condition and not as an assumption  \cite{Koerber:2007hd}. }
\begin{equation}
\label{eq:equalnorms}
 ||\eta^1_+||^2 = ||\eta^2_\mp||^2 \ , \qquad (\eta^1_+)^2 = (\eta^2_{\mp})^2 \ ,
\end{equation}
we see that $K_2$ takes exactly the same expression of $K_1$. Therefore we conclude that $K$ and $\tilde K$ are
\begin{equation}
\label{eq:KKtildefactorized}
 K= \frac{e^{-A}}{8} \bigl(z ||\eta^1_+||^2 + \mathrm{Re} (a (\eta^1_+)^2 ) \bigr) \ , \qquad \tilde K = 0 \ .
\end{equation}
It remains to consider the factorization of the fluxes and of the NSNS three-form $H$. The request of maximal symmetry in two dimensions imposes that all these fields (and also the dilaton) do not depend on the external coordinates $x^\mu$. Moreover the indices structure of them must be of the form
\begin{align}
\label{eq:fluxesdecomposition}
  & H = H_0 + H_2 \ , \nonumber \\
  & F = F_0 + F_2 = f + e^{2A} \mathrm{vol}_2 \wedge \ast_8 \lambda (f) \ ,
\end{align}
where the indices indicate the number of external components, $f$ is an internal polyform and the self-duality of $F$ (equation (\ref{eq:F*F})) is used. We can now move to discuss the system of equations (\ref{eq:susy10}) for these particular vacua.

\subsection{Symmetry equations}
\label{sec:symmetrygeneral}

To begin we consider the symmetry equations, i.e. the equations (\ref{eq:LK}). The first equation require that $K$ would be a Killing vector, however $K$ takes the expression (\ref{eq:KKtildefactorized}) and we already know that $z$ and $a$ are Killing vectors by construction (they are constant), therefore we obtain the constraints
\begin{align}
\label{eq:constraintsnorms}
  & ||\eta^1_+||^2 = \alpha e^A \ , \nonumber \\
  & (\eta^1_+)^2 = (\beta + i \delta) e^A \ ,
\end{align}
where $\alpha$, $\beta$ and $\delta$ are real constants. Moving to the second equation in (\ref{eq:LK}) it is straightforward to see (using (\ref{eq:KKtildefactorized})) that this equation implies
\begin{equation}
\label{eq:H2vanish}
  H_2 = 0 \ ,
\end{equation}
therefore in the following we will write $H$ to simply indicate $H_0$.

\subsection{Exterior equation}
\label{sec:exteriorgeneral}
We turn now to discuss the exterior equation (\ref{eq:psp10}) that, as remarked at the end of section {\ref{sec:SUSY10d}, in some situations contains all the information that we need.

We start by evaluating the r.h.s. in (\ref{eq:psp10}); it reads
\begin{equation}
\label{eq:psp10rhs}
- (\tilde K \wedge + \iota_K) F = \frac 18 (\alpha + \beta \mathrm{Re} (g) - \delta \mathrm{Im} (g)) e^{2a} z \wedge \ast_8 \lambda (f) \ , 
\end{equation}
where we used (\ref{eq:KKtildefactorized}), (\ref{eq:fluxesdecomposition}), (\ref{eq:constraintsnorms}), (\ref{eq:zpropa}), the self-duality of $z$ and the relation (valid for any $d$ even)
\begin{equation}
 \ast \lambda (d x^\mu \wedge) = - \iota_\mu \ast \lambda \ .
\end{equation}
Therefore, using the expression (\ref{eq:Phifactorized}) for the polyform $\Phi$, equation (\ref{eq:psp10}) becomes
\begin{equation}
 d_H \bigl(e^{A-\phi} \mathrm{Re} (z\wedge \psi_1 + a \wedge \psi_2)\bigr) = \mp \frac{1}{16} (\alpha + \beta \mathrm{Re} (g) - \delta \mathrm{Im} (g)) e^{2A} z \wedge \ast_8 \lambda (f) \ , 
\end{equation}
that can be decomposed in the couple of equations
\begin{align}
\label{eq:exteriordecomposed}
  & d_H (e^{A-\phi} \mathrm{Re} \psi_1) = \pm \frac{\alpha}{16} e^{2A} \ast_8 \lambda (f) \ , \nonumber \\
  & d_H (e^{A-\phi} \psi_2) = \pm \frac{\beta + i \delta}{16} e^{2A} \ast_8 \lambda (f) \ .
\end{align}

\subsection{Pairing equations}
\label{sec:pairinggeneral}

It remains to consider the pairing equations (\ref{eq:++1}) and (\ref{eq:++2}). We will present the computation only for (\ref{eq:++1}) since (\ref{eq:++2}) is completely parallel. The first part of the analysis will be very similar to the corresponding one presented in \cite{Tomasiello:2011eb} for four-dimensional vacua and therefore we will be brief. 

To start with, we have to choose the vectors $e_{+1}$ and $e_{+2}$. Since we have $K_1 = K_2 = K$ we can take $e_{+1} = e_{+2} = e_+$ as well, moreover we take $e_+$ purely external as $K$ and the action of the gamma matrices $\stackrel \rightarrow{\gamma}_+$ and $\stackrel \leftarrow{\gamma}_+$ takes the form
\begin{equation}
\label{eq:gamma_+action}
  \stackrel \rightarrow{\gamma}_+ = e^A e_+ \wedge + e^{-A} e_+ \llcorner \ , \qquad \stackrel \leftarrow{\gamma}_+ (-)^{\mathrm{deg}} = e^A e_+ \wedge - e^{-A} e_+ \llcorner \ .
\end{equation}

Now we can compute the various terms appearing in (\ref{eq:++1}): since $e_+$ is purely external the term containing $d^\dagger (e^{-2\phi} e_+)$ vanishes, moreover the term $d_H (e^{-\phi} \Phi \cdot e_+)$ can be massaged using
\begin{equation}
 \left\{ d , \stackrel \leftarrow{\gamma}_+ (-)^{\mathrm{deg}} \right\} = e^{-A} \partial_+ + dA \wedge \stackrel \rightarrow{\gamma}_+  \ .
\end{equation}
Summarizing, (\ref{eq:++1}) becomes
\begin{equation}
\label{eq:++1massaged}
\bigl(\gamma_+ \cdot \Phi \cdot \gamma_+, \Gamma^{MN} \bigl[dA \wedge \gamma_+ e^{-\phi} \Phi - 2 f \bigr]\bigr) = 0 \ ,
\end{equation}
where we used (\ref{eq:F*F}) and (\ref{eq:fluxesdecomposition}). 

We have now to evaluate (\ref{eq:++1massaged}) for the various possible choices of the indices $M$ and $N$. It is straightforward to see that for $M$ and $N$ both internal or external the equation reduces to an identity and so it has no content. Therefore the only non trivial equations come when we have $M=m$ and $N=\mu$. We start by computing the factor
\begin{equation}
\label{eq:++1secondaddone}
  - 2 \bigl(\gamma_+ \cdot \Phi \cdot \gamma_+, \Gamma^m \Gamma^\mu f \bigr) = \pm \frac{1}{16} \bar{\epsilon}_1 \gamma_+ \Gamma^m \Gamma^{\mu} f \gamma_+ \epsilon_2 \ ,
\end{equation}
where we used the identity
\begin{equation}
\label{eq:breakpairing}
  \bigl(\gamma_{+1} \cdot \Phi \cdot \gamma_{+2}, C \bigr) = - \frac{(-)^{\mathrm{deg} \Phi}}{32} \bar \epsilon_1 \gamma_{+1} C \gamma_{+2} \epsilon_2 \ ,
\end{equation}
that can be found in \cite{Tomasiello:2011eb}. Using now the equations (\ref{eq:spinans}) and (\ref{eq:gammadec}), we can further massage (\ref{eq:++1secondaddone}) obtaining
\begin{equation}
\label{eq:++1secondaddmassage}
- 2 \bigl(\gamma_+ \cdot \Phi \cdot \gamma_+, \Gamma^m \Gamma^\mu f \bigr) = \frac{1}{16} \bigl(\bar \zeta \gamma_+ \gamma^\mu \gamma_+ \zeta\,\eta^{1 \dagger}_+ \gamma^m f \eta^2_\mp + \bar \zeta \gamma_+ \gamma^\mu \gamma_+ \zeta^c \,\eta^{1 \dagger}_+ \gamma^m f \eta^{2 c}_\mp + \mathrm{c.c.}   \bigr) \ ,
\end{equation}
where the reality of the gamma matrices $\gamma^m$ and $\gamma^\mu$ was used.

A similar treatment can be reserved to the other term in (\ref{eq:++1massaged}) which finally takes the form
\begin{equation}
\label{eq:++1firstaddmassage}
  e^{-\phi} \bigl(\gamma_+ \cdot \Phi \cdot \gamma_+, \Gamma^m \Gamma^\mu [dA \wedge \gamma_+ \Phi] \bigr) = \pm \frac{e^{-\phi}}{4} \bigl(\bar \zeta \gamma_+ \gamma^\mu \gamma_+ \zeta\,\eta^{1 \dagger}_+ \gamma^m \partial A \eta^2_\mp + \bar \zeta \gamma_+ \gamma^\mu \gamma_+ \zeta^c \,\eta^{1 \dagger}_+ \gamma^m \partial A \eta^{2 c}_\mp + \mathrm{c.c.}   \bigr) \ .
\end{equation} 

To proceed further we observe that the two-dimensional bilinears take the form
\begin{align}
\label{eq:pairing4dpart}
  &\bar \zeta \gamma_+ \gamma^\mu \gamma_+ \zeta \propto e_+^\mu \ , \nonumber \\
  & \bar \zeta \gamma_+ \gamma^\mu \gamma_+ \zeta^c \propto \bar g e_+^\mu \ ,
\end{align}
therefore, requiring that (\ref{eq:++1massaged}) has a solution which is independent from $\zeta$, we conclude that (\ref{eq:++1secondaddmassage}) and (\ref{eq:++1firstaddmassage}) give rise to the following equations
\begin{align}
\label{eq:++1traces}
  & \mathrm{Re} (4 \eta^{1 \dagger}_+ \gamma^m \partial A \eta^1_+ \pm e^\phi \eta^{1 \dagger}_+ \gamma^m f \eta^{2}_\mp) = 0 \ , \nonumber \\
  & 4 \eta^{1 \dagger}_+ \gamma^m \partial A \eta^{1 c}_+ \pm e^\phi \eta^{1 \dagger}_+ \gamma^m f \eta^{2 c}_\mp = 0 \ ,
\end{align}
that can be recast in a more familiar fashion
\begin{align}
\label{eq:++1tracesfinal}
  & \mathrm{Re} \mathrm{Tr} \left(\eta^2_\mp \eta^{1 \dagger}_+ \gamma^m \left(4 \partial A \frac{\eta^1_+ \eta^{2 \dagger}_\mp}{||\eta^2_\mp||^2} \pm e^\phi f\right) \right) = 0 \ , \nonumber \\
  & \mathrm{Tr} \left(\eta^{2 c}_\mp \eta^{1 \dagger}_+ \gamma^m \left(4 \partial A \frac{\eta^{1 c}_+ \eta^{2 c \dagger}_\mp}{||\eta^2_\mp||^2} \pm e^\phi f\right) \right) = 0 \ ,
\end{align}
or, in terms of the eight-dimensional Chevalley-Mukai pairing, as
\begin{align}
\label{eq:++1mukai}
  & \mathrm{Re} \bigl(\gamma^m \bar \psi_1, dA \wedge \psi_1 \mp \frac{\alpha}{8} e^{\phi + A} \ast_8 \lambda (f) \bigr) = 0 \ , \nonumber \\
  & \bigl(\gamma^m \bar \psi_2, dA \wedge \bar \psi_1 \mp \frac{\alpha}{8} e^{\phi + A} \ast_8 \lambda (f) \bigr) = 0 \ .
\end{align}
Finally, equation (\ref{eq:++2}) can be treated in the same way and the final result is
\begin{align}
\label{eq:++2mukai}
  & \mathrm{Re} \bigl(\bar \psi_1 \gamma^m, dA \wedge \psi_1 \mp \frac{\alpha}{8} e^{\phi + A} \ast_8 \lambda (f) \bigr) = 0 \ , \nonumber \\
  & \bigl( \bar \psi_2 \gamma^m, dA \wedge \psi_1 \mp \frac{\alpha}{8} e^{\phi + A} \ast_8 \lambda (f) \bigr) = 0 \ .
\end{align}

\subsection{Summary}
\label{sec:summarygeneralSUSY}
We have rewritten the conditions for unbroken supersymmetry (equations (\ref{eq:susy10})) for a $\mathrm{Mink}_2$, $(2,0)$ vacuum solution. The resulting system of equations is given by (\ref{eq:constraintsnorms}), (\ref{eq:H2vanish}), (\ref{eq:exteriordecomposed}), (\ref{eq:++1mukai}) and (\ref{eq:++2mukai}). Unfortunately equations (\ref{eq:++1mukai}) and (\ref{eq:++2mukai}) are not as elegant as (\ref{eq:exteriordecomposed}) and this is a typical feature of the system (\ref{eq:susy10}). However we will see that, assuming that the structure group of $\mathcal{M}_8$ is $SU(4)\times SU(4)$, the pairing equations can be recast in a concise and elegant form.

\section{Supersymmetry conditions: the pure case}
\label{sec:pure}

In this section we will see (motivated by the results found in \cite{Prins:2013koa}) how SUSY conditions can be rewritten in a compact form if we make the assumption that the internal spinors $\eta^1_+$ and $\eta^2_\mp$ are {\it pure}.\footnote{Recall that a spinor is said to be pure if it is annihilated by exactly half of the gamma matrices.} In this case it is possible to show that the structure group of the generalized tangent bundle $T_8 \oplus T^\ast_8$ reduces to $SU(4) \times SU(4)$ and this allows a better formulation of the pairing equations. The equations that we will find are already present in \cite{Prins:2013koa} but, contrary to that work, we will not assume that the two spinors $\eta^1_+$ and $\eta^2_\mp$ are proportional (notice that such an assumption can be done in Type IIB only). Therefore our results in this section can be seen as the extension from the {\it strict} $SU(4)$ case (treated in \cite{Prins:2013koa}) to the $SU(4) \times SU(4)$ case.

\subsection{Pure spinors and generalized Hodge diamonds}
\label{sec:generalizedhodgepure}
Let us start by reviewing what the purity condition on eight-dimensional spinors implies and what geometrical structures can be defined on $\mathcal{M}_8$ when $\eta^1_+$ and $\eta^2_\mp$ are pure.

Contrary to what happens in lower dimensions, in eight dimensions Weyl spinors are not necessarily pure, as shown by a simple counting argument: in eight dimensions the space of pure spinors has real dimension $14$ whereas the space of Weyl spinors has real dimension $16$. More explicitly, a given eight-dimensional Weyl spinor of (say) positive chirality $\eta_+$ is pure if and only if it satisfies the additional algebraic condition
\begin{equation}
\label{eq:puritycondition}
    \eta^t_+ \eta_+ = 0 \ .
\end{equation}
Notice that a Majorana-Weyl spinor cannot be pure. In this section we will suppose that both $\eta^1_+$ and $\eta^2_\mp$ satisfy (\ref{eq:puritycondition}) and hence that they are pure.

In general, it is well known that a pure spinor in $2d$ dimensions implies that the structure group on the tangent bundle reduces to $SU(d)$. This is equivalent to saying that on the manifold a real two-form $J$ and a $(d,0)$-form (with respect to the almost complex structure defined by $J$) called $\Omega_d$ can be defined. $J$ and $\Omega_d$ can be obtained from $\eta_+$ via the relations
\begin{equation}
\label{eq:spinorsforms}
  i J_{mn} = \eta^\dagger_+ \gamma_{mn} \eta_+ \ , \qquad \Omega_{m_1 \dots m_d} = \eta^t_+ \gamma_{m_1 \dots m_d} \eta_+ \ ,
\end{equation}
and they satisfy
\begin{equation}
\label{eq:SU(d)structurerel}
\frac{1}{2^d} \Omega_d \wedge \bar \Omega_d = \frac{1}{d!} J^d = \mathrm{vol}_{2d} \ , \qquad J \wedge \Omega_d = 0 \ , 
\end{equation}
that can be shown from (\ref{eq:spinorsforms}) by Fierzing. Having introduced $J$ we can reformulate the purity condition of $\eta_+$ by saying that it is annihilated by the gamma matrices holomorphic with respect to the almost complex structure defined by $J$.

It is also known that the situation is much more involved when we consider a pair of pure spinors $\eta^1_+$ and $\eta^2_\mp$, indeed we have that the structure group depends on the chirality of the two spinors and also it can acquire a dependence on the points of the manifold. To treat all these different situations on the same footing it is often useful to consider the generalized tangent bundle $T \oplus T^\ast$, since in this enlarged space the structure group is always $SU(d) \times SU(d)$.

On $T\oplus T^\ast$ we can define a $Cl(2d,2d)$ algebra, with the corresponding gamma matrices given by
\begin{equation}
\label{eq:generalizedgamma}
\Gamma_\Lambda = \left\{\partial_1 \llcorner, \dots, \partial_{2d}, dx^1 \wedge, \dots, dx^{2d} \wedge \right\} \ ,
\end{equation}
and with spinors simply given by the differential forms of all degrees. Now the key point is the following: starting from the pure spinors $\eta^1_+$ and $\eta^2_\mp$ (which are pure spinors with respects to the usual $Cl(2d)$ algebra), we can define the polyforms $\psi_1$ and $\psi_2$ that we defined in (\ref{eq:psidefinition}) and, thanks to the purity of $\eta^1_+$ and $\eta^2_\mp$, we are sure that they are pure with respects to the $Cl(2d,2d)$ algebra defined in (\ref{eq:generalizedgamma}).

To any given pure spinor $\psi_i$ on the generalized tangent bundle one can  associate a {\it generalized almost complex structure} $\mathcal{J}_i$ (GACS), i.e. an operator $
\mathcal{J}_i: T \oplus T^\ast \rightarrow T \oplus T^\ast$ such that $\mathcal{J}_i^2 = - 1$; the relation between $\psi_i$ and $\mathcal{J}_i$ is given by the requirement that the $i$-eigenbundle of $\mathcal{J}_i$ coincides with the annihilator of $\psi_i$. 

Finally, it can be shown that $\psi_1$ and $\psi_2$ constructed as bilinears of $\eta^1_+$ and $\eta^2_\mp$ are {\it compatible} which means that the corresponding GACSs commute.

Specializing now the discussion to the eight-dimensional case, we want to introduce an appropriate basis for the differential forms on $\mathcal M_8$. To this end it is useful to consider the so-called generalized Hodge diamond, which constitutes a basis for the differential forms of any degrees constructed starting from $\psi_1$ and $\psi_2$. We can represent this basis as follows:
\begin{equation}
  \label{eq:hodge}
  \begin{array}{c}\vspace{.1cm}
\psi_1 \\ \vspace{.1cm}
\psi_1\gamma^{i_2}  \hspace{1cm} \gamma^{\bar i_1}\psi_1 \\
\psi_1\gamma^{i_2j_2} \hspace{1cm} \gamma^{\bar i_1} \psi_1\gamma^{i_2}
\hspace{1cm} \gamma^{\bar i_1 \bar j_1} \psi_1\\
\psi_2\gamma^{\bar i_2} \hspace{1.2cm}\gamma^{\bar i_1}\psi_1\gamma^{i_2 j_2}
\hspace{1cm} \gamma^{\bar i_1 \bar  j_1}\psi_1\gamma^{j_2}
\hspace{1.2cm}\gamma^{i_1}\bar \psi_2\\
 \psi_2 \hspace{1.2cm} \gamma^{\bar i_1}\psi_2\gamma^{\bar i_2} \hspace{1.2cm} \gamma^{\bar i_1\bar j_1}\psi_1\gamma^{i_2j_2} \hspace{1.2cm} \gamma^{\bar i_1} \bar \psi_2 \gamma^{i_2} \hspace{1.2cm}\bar \psi_2 \\
 \gamma^{\bar i_1} \psi_2 \hspace{1.2cm} \gamma^{i_1 j_1}\bar\psi_1\gamma^{\bar i_2} \hspace{1cm} \gamma^{i_1}\bar\psi_1\gamma^{\bar i_2 \bar j_2} \hspace{1.2cm}\bar\psi_2\gamma^{i_2}\\
 \gamma^{i_1 j_1} \bar\psi_1 \hspace{1cm} \gamma^{i_1}\bar\psi_1\gamma^{\bar i_2} \hspace{1cm}\bar\psi_1\gamma^{\bar i_2 \bar j_2} \\
 \gamma^{i_1}\bar\psi_1 \hspace{1cm} \bar\psi_1\gamma^{\bar i_2} \\ \vspace{.1cm}
 \bar\psi_1
  \end{array}\
\end{equation}
where the action of the gamma matrices on $\psi_i$ is obviously obtained from the same action on the spinors $\eta^i$. 

This basis has the property of being {\it orthogonal}: every form has vanishing Chevalley-Mukai pairing with every form in the diamond, except with the ones symmetric with respect to the central point. So for example $\psi_1$ has non vanishing pairing only with $\bar\psi_1$, $\psi_1 \gamma^{i_2}$ only with $\bar\psi_1 \gamma^{\bar i_2}$ and so on. Another important technical property of this basis is that its entries are eigenfunctions for the action of $(\mathcal{J}_1 \cdot, \mathcal{J}_2 \cdot)$ corresponding to $\psi_1$ and $\psi_2$, and also for the operator $\ast_8 \lambda$. More explicitly, the eigenvalues for all these operators are
\begin{equation}\label{eq:eigenvalues}
	\begin{picture}(100,100)(180,0)
		\put(-10,60){$({\cal J}_1 \cdot\,,{\cal J}_2\cdot)\ :$}
		\put(0,0){$
\begin{array}{c}
		(4i,0)\\
		(3i,i) \hspace{.7cm} (3i,-i) \\ 
		(2i,2i) \hspace{.7cm}(2i,0) 
		\hspace{.7cm} (2i,-2i)\\
		(i,3i) \hspace{.7cm} (i,i)
		\hspace{.7cm} (i,-i)
		\hspace{.7cm}(i,-3i)\\
		(0,4i) \hspace{.7cm} (0,2i) \hspace{.9cm}(0,0) \hspace{.7cm}(0,-2i)\hspace{.7cm}(0,-4i) \\
		(-i,3i) \hspace{.9cm} (-i,i) \hspace{.7cm}(-i,-i) \hspace{.7cm}(-i,-3i) \\
		 (-2i,2i) \hspace{.7cm} (-2i,0)
		\hspace{.7cm}(-2i,-2i)\\
		(-3i,i) \hspace{.7cm} (-3i,-i)\\
		 (-4i,0)\\
	  \end{array}\\$} 
	 	\put(270,60){$\ast_8\lambda\ :$}
	 	\put(280,0){$
	\begin{array}{c}\vspace{.1cm}
		+\\
		+  \hspace{.5cm} - \\ 
		+ \hspace{.5cm}- 
		\hspace{.5cm} +\\
		+ \hspace{.5cm} -
		\hspace{.5cm} +
		\hspace{.5cm}-\\
		+ \hspace{.5cm} - \hspace{.5cm}+ \hspace{.5cm}-\hspace{.5cm}+ \\
		- \hspace{.5cm} + \hspace{.5cm}- \hspace{.5cm}+ \\
		 + \hspace{.5cm} -
		\hspace{.5cm}+\\
		- \hspace{.5cm} +\\
		 +\\
	  \end{array}\ $ }
		\put(420,-40){.}
			\end{picture}\vspace{2cm}
\end{equation}

\subsection{Rewriting SUSY conditions in the pure case}
\label{sec:susypure}
We have now all the instruments necessary to massage the system of equations found in section \ref{sec:SUSYgeneral} with the assumption that $\eta^1_+$ and $\eta^2_\mp$ are pure.

First of all, to stay closer to the results of \cite{Prins:2013koa},  we perform the following redefinitions:
\begin{equation}
\label{eq:redefinitionsphi}
  \psi_1 = \frac{1}{e^A} \eta^1_+\eta^{2 \dagger}_\mp \ , \qquad \psi_2 = \frac{1}{e^A} \eta^1_+ \eta^{2  c \dagger}_\mp \ .
\end{equation}
Next we move to the symmetry equations (\ref{eq:constraintsnorms}): it is straightforward to see that the second equation implies 
\begin{equation}
\beta= \delta = 0 \ ,
\end{equation} 
since $\eta^1_+$ and $\eta^2_\mp$ are pure. Therefore we can interpret the geometrical role of $\beta$ and $\delta$ as parametrizing the departure from the purity condition. We will discuss this last statement in a more geometrical language in section \ref{sec:impure}.

Moving to the exterior equations (\ref{eq:exteriordecomposed}), taking into account the redefinition (\ref{eq:redefinitionsphi}) and the vanishing of $\beta$ and $\delta$, they become
\begin{align}
\label{eq:exteriorpure}
  & d_{H} (e^{2A-\phi} \mathrm{Re} \psi_1) = \pm \frac {\alpha}{16} e^{2A} \ast_8 \lambda (f) \ , \nonumber \\
  & d_{H} (e^{2A -\phi} \psi_2) = 0 \ .
\end{align}

It remains to consider the pairing equations. To start with we see that, using the orthogonality of the generalized Hodge diamond, the second equation in (\ref{eq:++1mukai}) can be simplified
\begin{equation}
  (\gamma^{i_1} \bar \psi_2, f) = 0 \ ,
\end{equation}
 and analogously the second equation in (\ref{eq:++2mukai}) becomes
\begin{equation}
  (\bar \psi_2 \gamma^{i_2}, f) = 0 \ .
\end{equation}
Collecting the results we have the following expression for the pairing equations\footnote{Notice that we have removed the real part in front of the first equations in (\ref{eq:++1mukai}) and (\ref{eq:++2mukai}). This is due to the fact that now the holomorphic (or anti-holomorphic) gamma matrices appear.}
\begin{align}
\label{eq:pairingpure}
  & \bigl(\gamma^{i_1} \bar \psi_1, dA \wedge \psi_1 \mp \frac{\alpha}{8} e^{\phi} \ast_8 \lambda (f) \bigr) = 0 \ , \nonumber \\
  & (\gamma^{i_1}\bar \psi_2, f) = 0 \ , \nonumber \\
  & \bigl(\bar \psi_1 \gamma^{\bar i_2}, dA \wedge \psi_1 \mp \frac{\alpha}{8} e^{\phi} \ast_8 \lambda (f) \bigr) = 0 \ , \nonumber \\
  & (\bar \psi_2 \gamma^{i_2}, f) = 0 \ .
\end{align}

By a direct computation, using the properties contained in (\ref{eq:eigenvalues}), it can be shown that the equations in (\ref{eq:pairingpure}) are equivalent to the single equation
\begin{equation}
\label{eq:pairingpurerewritten}
   d_{H}^{\mathcal{J}_2} (e^{-\phi} \mathrm{Im} \psi_1) = \pm \frac{\alpha}{16} f \ ,
\end{equation}
where $d_{H}^{\mathcal{J}_2} \equiv [d_{H}, \mathcal{J}_2 \cdot]$. The equivalence between (\ref{eq:pairingpure}) and (\ref{eq:pairingpurerewritten}) is in appendix \ref{sec:appendixpure}. 

\subsection{Summary}
\label{sec:summarypure}
Let us summarize the results of this section. We have shown that, assuming the purity of the spinorial parameters $\eta^1_+$ and $\eta^2_\mp$, SUSY equations can be reformulated in terms of three conditions
\begin{align}
	\label{eq:susypure}
\fbox{$
\begin{array}{rl}
		& d_{H} (e^{2A-\phi} \mathrm{Re} \psi_1) = \pm \frac {\alpha}{16} e^{2A} \ast_8 \lambda (f)\ ,\\
		& d_{H} (e^{2A -\phi} \psi_2) = 0\ ,\\
		& d_{H}^{\mathcal{J}_2} (e^{-\phi} \mathrm{Im} \psi_1) = \pm \frac{\alpha}{16} f\ .
\end{array}$}	
\end{align}
These equations were already found in \cite{Prins:2013koa} under the simplifying hypothesis of strict $SU(4)$ structure (and so only Type IIB theory was considered in that work). Therefore we have shown in this section that the results of \cite{Prins:2013koa} can be extended to the more general situation in which the SUSY parameters are {\it not} proportional, and this allows to treat Type IIA and Type IIB on the same footing. 

As already emphasized in the Introduction, our result is in perfect agreement with the results of \cite{Martucci:2011dn}: in that work it is shown that the calibrations issues involve only the symmetry equations (\ref{eq:LK}) and the exterior equation (\ref{eq:psp10}). On the other hand the pairing equations (\ref{eq:++1}) and (\ref{eq:++2}) have no counterpart in the calibrations recipe and indeed we find that the additional equation (\ref{eq:pairingpurerewritten}) is given exactly by the pairing equations. 


Of course it would be interesting to look for a generalization of the supersymmetry-calibrations correspondence which takes into account the pairing equations. Obtaining such a correspondence could give a more geometrical understanding of the pairing equations and perhaps a better formulation for them.

\section{Beyond the pure case}
\label{sec:impure}

In this section we will remove the hypothesis that $\eta^1_+$ and $\eta^2_\mp$ are pure spinors on $\mathcal{M}_8$. Nevertheless we will assume that a pair of pure spinors $\tilde \eta^1_+$ and $\tilde \eta^2_\mp$ on $\mathcal M_8$ exists. In other words we will assume that the structure group of the generalized tangent bundle on $\mathcal M_8$ is still $SU(4)\times SU(4)$ but the SUSY parameters $\eta^1_+$ and $\eta^2_\mp$ are {\it not} the spinors realizing the reduction of the structure group.  It will become clear in section \ref{sec:parametrization} that, at least locally, given a Weyl spinor $\eta$ one can always obtain a corresponding pure spinor $\tilde \eta$, by simply taking its real and imaginary parts and by rescaling them; however globally some obstructions can occur. In this section we will assume that such global obstructions do not occur and that we can find a pair of globally defined pure spinors.

\subsection{Parametrization of non-pure spinors}
\label{sec:parametrization}

Given the assumption that a pair of pure spinors on $\mathcal{M}_8$ exists we want to determine a parametrization of $\eta^1_+$ and $\eta^2_{\mp}$ in terms of the pure spinors $\tilde{\eta}^1_+$ and $\tilde{\eta}^2_\mp$.

To this end we start by recalling that a Weyl spinor (not Majorana) $\eta$\footnote{We have not written the chirality of $\eta$ since the discussion does not depend on it.} can be written in terms of two Majorana-Weyl spinors $\chi_1$ and $\chi_2$ as follows
\begin{equation}
\label{eq:MMW}
  \eta= \chi_1 + i \chi_2 \ .
\end{equation}  
(\ref{eq:MMW}) gives us a simple geometrical interpretation of the purity condition (\ref{eq:puritycondition}) as an orthonormality property of the spinors $\chi_1$ and $\chi_2$: indeed it is straightforward to see that $\eta$ is pure if and only if $\chi_1$ and $\chi_2$ satisfy
\begin{equation}
\label{eq:purityconditionMW}
  \chi^{ t}_1\chi_1 = \chi^{ t}_2 \chi_2 \ , \qquad \chi^{ t}_1 \chi_2 = 0 \ .
\end{equation}
In other words, a Weyl spinor $\eta$ is pure if and only if its Majorana-Weyl components $\chi_1$ and $\chi_2$ have the same norms (the first condition in (\ref{eq:purityconditionMW})) and they are orthogonal (the second condition in (\ref{eq:purityconditionMW})). On the other hand, we see that the obstacles to the purity of $\eta$ are given by a difference in the norms of $\chi_1$ and $\chi_2$ or if they are not orthogonal.

To proceed, suppose that we have, beyond the non-pure spinor $\eta$, a pure spinor $\tilde{\eta}$ with the same chirality and with components $\tilde{\chi}_1$ and $\tilde{\chi}_2$. For future convenience we take the norms of $\tilde{\chi}_1$ and $\tilde{\chi}_2$ to be equal to $e^{A(y)}$ (where $A(y)$ is of course the warping factor appearing in (\ref{eq:metric}))
\begin{equation}
\label{eq:puresupportdefinition}
  \tilde{\chi}^{ t}_1\tilde{\chi}_1 = \tilde{\chi}^{ t}_2\tilde{\chi}_2 = e^{A(y)} \quad \Rightarrow \quad ||\tilde{\eta}||^2 = 2 e^{A(y)} \ ,
\end{equation}
we also apply a rotation to $\tilde{\eta}$ in order to put $\tilde{\chi}_1$ along $\chi_1$. A pictorial description of this construction is given in figure \ref{fig:nonpureparam} which shows that $\eta$ can be parametrized in terms of $\tilde{\eta}$ (and its complex conjugate) via the formula
\begin{equation}
\label{eq:parametrizationeta}
  2 \eta = \bigl(A_1 + i B_1 e^{-i \theta_1} \bigr) \tilde{\eta} + \bigl(A_1 + i B_1 e^{i\theta_1} \bigr) \tilde{\eta}^c \ ,
\end{equation}
where the real quantities $A_1$ and $B_1$ are given by
\begin{equation}
A_1 = \sqrt{\frac{\chi^{ t}_1 \chi_1}{e^{A(y)}}} \ , \qquad B_1 =\sqrt{ \frac{\chi^{ t}_2 \chi_2}{e^{A(y)}}} \ ,
\end{equation}
and $\theta_1$ parametrizes the angle between $\chi_1$ and $\chi_2$.

As a check of the validity of this parametrization notice that $\tilde{\eta}$ is a pure spinor of fixed norm, hence it has $13$ real components; on the other hand $A_1$, $B_1$ and $\theta_1$ are  real coefficients. This gives us a total of $16$ real components for $\eta$ which is correct for a Weyl non-pure spinor on $\mathcal{M}_8$. We note also that in the pure limit we have $A_1=B_1$ and $\theta_1=\frac{\pi}{2}$ for a total of $14$ real components as it should.

\begin{figure}[ht]
	\centering
		\includegraphics[scale=1]{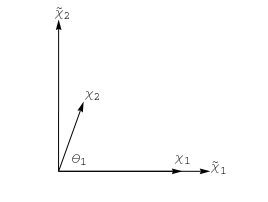}
	\caption{A pictorial description of the parametrization (\ref{eq:parametrizationeta}). The Majorana-Weyl components of the non-pure spinor $\eta$ are $\chi_1$ and $\chi_2$; they can be represented as a couple of vectors with different norms and forming an angle $\theta_1$. On the other hand the Majorana-Weyl components of the pure spinor $\tilde{\eta}$ are given by $\tilde{\chi}_1$ and $\tilde{\chi}_2$; they have the same norm $\tilde{\chi}^{ t}_1 \tilde{\chi}_1 = \tilde{\chi}^{ t}_2  \tilde{\chi}_2 = e^{A(y)}$ and they are orthogonal. $A_1$ and $B_1$ appearing in (\ref{eq:parametrizationeta}) are given by $A_1 = \sqrt{\frac{\chi^{ t}_1 \chi_1}{e^{A(y)}}}$, $B_1 =\sqrt{ \frac{\chi^{ t}_2 \chi_2}{e^{A(y)}}}$.}
	\label{fig:nonpureparam}
\end{figure}

These considerations can be applied to the SUSY parameters $\eta^1_+$ and $\eta^2_\mp$ which in terms of the pure spinors $\tilde{\eta}^1_+$ and $\tilde{\eta}^2_\mp$ read
\begin{align}
\label{eq:SUSYparametrization}
  & 2 \eta^{1}_+ = c_1 \tilde{\eta}^1_+ + c_2 \tilde{\eta}^{1 c}_+ \ , \nonumber \\
  & 2 \eta^2_\mp = c_3 \tilde{\eta}^2_\mp + c_4 \tilde{\eta}^{2 c}_\mp \ ,
\end{align} 
where
\begin{align}
\label{eq:fidefinition}
  & c_1 \equiv A_1 + i B_1 e^{- i \theta_1} \ , \qquad c_2 \equiv A_1 + i B_1 e^{ i \theta_1} \ , \nonumber \\
  & c_3 \equiv A_2 + i B_2 e^{- i \theta_2} \ , \qquad c_4 \equiv A_2 + i B_2 e^{ i \theta_2} \ ,
\end{align}
we will see in a moment that all these coefficients are {\it constant} on $\mathcal{M}_8$.
Thanks to the parametrization (\ref{eq:SUSYparametrization}) we can now massage the conditions for unbroken supersymmetry deduced in section \ref{sec:SUSYgeneral}.

\subsection{Symmetry equations}
\label{sec:symmetryimpure}

We start by massaging the symmetry equations that we already wrote in full generality in (\ref{eq:constraintsnorms}). Putting (\ref{eq:SUSYparametrization}) in (\ref{eq:constraintsnorms}) and using the assumption that $||\tilde{\eta}^1_+||^2 = ||\tilde{\eta}^2_\mp||^2 = 2 e^A$ we obtain, after some manipulations, the equations
\begin{align}
\label{eq:ficonstants}
  & \beta + i \delta = c_1c_2 \ , \qquad \beta + i \delta = c_3 c_4 \ , \nonumber \\
  & 2\alpha = |c_1|^2 + |c_2|^2 \ , \qquad 2\alpha = |c_3|^2 + |c_4|^2 \ . 
\end{align} 
If we recall the definitions of the coefficients $c_i$ given in (\ref{eq:fidefinition}), we see that (\ref{eq:ficonstants}) leads to
\begin{equation}
\label{eq:alphabetadelta}
\alpha = A_{1,2}^2 + B_{1,2}^2 \ , \qquad \beta = A_{1,2}^2 - B_{1,2}^2 \ , \qquad \delta =  2 A_{1,2} B_{1,2} \cos \theta_{1,2} \ ,
\end{equation}
which clarifies the geometrical interpretation of $\beta$ and $\delta$: they express the departure from the purity condition, $\beta$ parametrizes a difference in the norms of the Majorana-Weyl components, $\delta$ keeps into account a lacking of orthogonality.

As an immediate consequence of (\ref{eq:alphabetadelta}) we see that $c_1=c_3$, $c_2 = c_4$ and, more important, that they are {\it constant} as promised. We therefore rewrite (\ref{eq:SUSYparametrization}) as
\begin{align}
\label{eq:SUSYparametrizationone}
& 2 \eta^{1}_+ = c_1 \tilde{\eta}^1_+ + c_2 \tilde{\eta}^{1 c}_+ \ , \nonumber \\
  & 2 \eta^2_\mp = c_1 \tilde{\eta}^2_\mp + c_2 \tilde{\eta}^{2 c}_\mp \ .
\end{align}

\subsection{Exterior equations}
\label{eq:exteriorimpure}

Let us now consider the exterior equations (\ref{eq:exteriordecomposed}). Having introduced the pure spinors $\tilde{\eta}^1_+$ and $\tilde{\eta}^2_\mp$ we can use the parametrization (\ref{eq:SUSYparametrizationone}) to deduce an analogous parametrization of the bilinears $\psi_1$ and $\psi_2$ in terms of the pure spinors $\tilde{\psi}_1$ and $\tilde{\psi}_2$ constructed from $\tilde{\eta}^1_+$ and $\tilde{\eta}^2_\mp$:
\begin{align}
\label{eq:bilinearspureimpure}
  &\psi_1 = \frac 14 \bigl[|c_1|^2 \tilde{\psi}_1 + |c_2|^2 \bar{\tilde{\psi}}_1 + c_1 \bar{c}_2 \tilde{\psi}_2 + \bar{c}_1 c_2 \bar{\tilde{\psi}}_2 \bigr] \ , \nonumber \\
  & \psi_2 = \frac 14 \bigl[c_1c_2 (\tilde{\psi}_1 + \bar{\tilde{\psi}}_1) + c_1^2 \tilde{\psi_2} + c_2^2 \bar{\tilde{\psi}}_2 \bigr] \ ,
\end{align}
in the pure limit we have $\psi_1 = A_1^2 \tilde{\psi}_1$ and $\psi_2 = A_1^2 \tilde{\psi}_2$ as it should. (\ref{eq:bilinearspureimpure}) can be put into  (\ref{eq:exteriordecomposed}) that becomes
\begin{align}
\label{eq:exteriorimpureone}
  & 2\alpha\, d_H \bigl(e^{2A - \phi} \mathrm{Re} \tilde{\psi}_1 \bigr) + c_1 \bar{c}_2 d_H \bigl(e^{2A - \phi} \tilde{\psi}_2 \bigr) + \bar{c}_1 c_2 d_H \bigl(e^{2A - \phi} \bar{\tilde{\psi}}_2 \bigr) = \pm \frac{\alpha}{4} e^{2A} \ast_8 \lambda (f) \ , \nonumber \\
  & 2 c_1c_2 d_H \bigl(e^{2A - \phi} \mathrm{Re} \tilde{\psi}_1 \bigr) + c_1^2 d_H \bigl(e^{2A - \phi} \tilde{\psi}_2 \bigr) + c_2^2 d_H \bigl(e^{2A - \phi} \bar{\tilde{\psi}}_2 \bigr) = \pm \frac{c_1 c_2}{4} e^{2A} \ast_8 \lambda (f) \ ,
\end{align}
where we used $2 \alpha = |c_1|^2+|c_2|^2$. At first sight these equations are not as pleasant as one might wish however, by simply expressing the coefficients $c_1$, $c_2$ and $\alpha$ in terms of $A_1$, $B_1$ and $\theta_1$ as in (\ref{eq:fidefinition}) and (\ref{eq:alphabetadelta}), and by separating the real and the imaginary part in the second equation in (\ref{eq:exteriorimpureone}), it can be shown with some simple manipulations that they are equivalent to
\begin{align}
\label{eq:exteriorimpurefinal}
  & d_H (e^{2A-\phi} \mathrm{Re} \tilde{\psi}_1) = \pm \frac{1}{8} e^{2A} \ast_8 \lambda (f) \ , \nonumber \\
  & d_H (e^{2A -\phi}  \tilde{\psi}_2) =0 \  .
\end{align} 
Rewritten in this form the geometrical content of these equations is much more transparent: apart from the trivial redefinition $\tilde{\psi}_1 \to \frac \alpha2 \tilde{\psi}_1$ we see that (\ref{eq:exteriorimpurefinal}) take {\it exactly} the same form of the equations (\ref{eq:exteriorpure}) which are valid in the pure case. In other words, we have deduced that, given the assumption that the structure group on $T_8 \oplus T^\ast_8$ is $SU(4) \times SU(4)$, the exterior equations, when expressed in terms of pure spinors on the generalized tangent bundle, take {\it always} the same form, no matter whether the spinorial parameters $\eta^1_+$ and $\eta^2_\mp$ are pure or not. It is possible that a better understanding of such a behaviour can be obtained from the calibrations perspective.

\subsection{Pairing equations}
\label{sec:pairingimpure}

It remains to massage the pairing equations that, as usual, are much more intricate than the others. The strategy can be easily described: as we have seen in section \ref{sec:pure} and in appendix \ref{sec:appendixpure}, generalized complex geometry (and in particular the generalized Hodge diamond (\ref{eq:hodge}) and its properties (\ref{eq:eigenvalues})) gives us a way to rewrite the pairing equations in a fancy form when $\eta^1_+$ and $\eta^2_\mp$ are pure. It is therefore conceivable that a similar simplification arises also in the non-pure case, thanks to the generalized Hodge diamond constructed from $\tilde{\psi}_1$ and $\tilde{\psi}_2$. 

Given the strategy just described we show in appendix \ref{sec:appendiximpure} that the pairing equations (\ref{eq:++1mukai}), (\ref{eq:++2mukai}) can be rewritten in terms of the pure spinors $\tilde{\psi}_1$ and $\tilde{\psi}_2$ as\footnote{\label{foot:one}The indices $i_1$ and $i_2$ should be intended as $\tilde{i}_1$, $\tilde{i}_2$, meaning that we are taking holomorphic indices with respect to the almost complex structures defined by the pure spinors $\tilde{\eta}^1_+$ and $\tilde{\eta}^2_\mp$ respectively. However we will use the notations $i_1$ and $i_2$ just for simplicity.}
\begin{subequations}
\label{eq:pairingimpuremain}
\begin{align}
\label{eq:pairingimpuremainone}
 & \bigl(\gamma^{i_1} \bar{\tilde{\psi}}_1, dA \wedge \tilde{\psi}_1 \mp \frac 14 e^\phi \ast_8 \lambda (f) \bigr)= 0 \ , \nonumber \\
 & \bigl( \bar{\tilde{\psi}}_1 \gamma^{\bar i_2}, dA \wedge \tilde{\psi}_1 \mp \frac 14 e^\phi \ast_8 \lambda (f) \bigr)= 0 \ , 
\end{align}
and
\begin{align}
\label{eq:pairingimpuremaintwo}
 & \bigl (\gamma^{i_1} \bar{\tilde{\psi}}_2, \ast_8 \lambda (f) \bigr) \mp \frac{8 \bar c e ^\phi}{\alpha} \bigl(\gamma^{i_1} \bar{ \tilde{\psi}}_2, \frac{dA}{4} \wedge \tilde \psi_2 \bigr) = \pm \omega^{i_1} \  , \nonumber \\
 & \bigl (\bar{\tilde{\psi}}_2 \gamma^{i_2}, \ast_8 \lambda (f) \bigr) \mp \frac{8 \bar c e ^\phi}{\alpha} \bigl( \bar{ \tilde{\psi}}_2 \gamma^{i_2}, \frac{dA}{4} \wedge \tilde \psi_2 \bigr) = \pm \sigma^{i_2} \  ,
\end{align}
\end{subequations}
where we defined
\begin{align}
  & \omega^{i_1} \equiv \mp \frac{(b+ 2\alpha) \bar{d} e^{2\phi}}{2 \alpha \bar e} \bigl(\gamma^{i_1} \bar{\tilde{\psi}}_1, \ast_8 \lambda (f) \bigr) \ , \nonumber \\
  & \sigma^{i_2} \equiv \mp \frac{(b+ 2\alpha) \bar{d} e^{2\phi}}{2 \alpha \bar e} \bigl( \tilde{\psi}_1 \gamma^{i_2}, \ast_8 \lambda (f) \bigr) \ ,
\end{align}
and the quantities $b$, $c$, $d$, $e$ are defined in (\ref{eq:shortcuts}). We note that, apart from the trivial redefinition $\tilde{\psi}_1 \to \frac \alpha2 \tilde{\psi}_1$ already noted after (\ref{eq:exteriorimpurefinal}), (\ref{eq:pairingimpuremainone}) again reproduces the corresponding ones valid in the pure case (first and third equations in (\ref{eq:pairingpure})), on the other hand (\ref{eq:pairingimpuremaintwo}) are similar to the pure case (second and fourth equations in (\ref{eq:pairingpure})) but contain additional deformation pieces (that of course vanish in the pure limit). It is therefore natural to look for a formulation of (\ref{eq:pairingimpuremain}) which is similar to (\ref{eq:pairingpurerewritten}), and indeed, using the same techniques of appendix \ref{sec:appendixpure}, we can recast (\ref{eq:pairingimpuremain}) as
\begin{equation}
\label{eq:pairingimpuremainfinal}
 d_H^{\tilde{\mathcal{J}}_2} \bigl(e^{-\phi} \mathrm{Im} \tilde{\psi}_1 \bigr) = \pm \frac 18 f - \mathrm{Re} \bigl( \frac{2 \bar c e^\phi}{\alpha} \ast_8 \lambda (dA \wedge  \tilde{\psi}_2) \bigr) + \mathrm{Re} \bigl(\gamma^{\bar{i}_1} \tilde{\psi}_2 \hat{\omega}_{\bar{i}_1} \bigr) - \mathrm{Re} \bigl(\psi_2 \gamma^{\bar{i}_2} \hat{\sigma}_{\bar{i}_2} \bigr) \ ,
\end{equation}  
where we introduced
\begin{equation}
\label{eq:omegahat}
  \hat{\omega}_{\bar{i}_1} \equiv \delta_{\bar{i}_1 j_1} \bigl(\gamma^{j_1} \bar{\tilde{\psi}}_2 , \gamma^{\bar{l}_1} \tilde{\psi}_2 \bigr)^{-1} \delta_{\bar{l}_1 i_1} \omega^{i_1} \ , \qquad \hat{\sigma}_{\bar{i}_2} \equiv \delta_{\bar{i}_2 j_2} \bigl( \bar{\tilde{\psi}}_2 \gamma^{j_2}, \tilde{\psi}_2 \gamma^{\bar{l}_2} \bigr)^{-1} \delta_{\bar{l}_2 i_2}\sigma^{i_2} \ .
\end{equation}

\subsection{Summary}
\label{sec:impuresummary}
Summarizing the results of this section, we have removed the purity condition (\ref{eq:puritycondition}) on the spinorial parameters. Nevertheless we have assumed that a couple of pure spinors $\tilde{\eta}^1_+$ and $\tilde{\eta}^2_\mp$ on $\mathcal{M}_8$ exists and in this way we have obtained the parametrization (\ref{eq:SUSYparametrization}). The conditions for unbroken supersymmetry enforce the coefficients of this parametrization to be {\it constant} on $\mathcal{M}_8$. Moreover we have shown that the exterior equations (\ref{eq:exteriordecomposed}), when rewritten in terms of the pure spinors $\tilde{\psi_1}$ and $\tilde{\psi}_2$, take exactly the same form of the pure case (\ref{eq:exteriorpure}). On the other hand the pairing equations (\ref{eq:++1}), (\ref{eq:++2}) are different but nevertheless can be recast in an elegant form (\ref{eq:pairingimpuremainfinal}) which can be interpreted as a deformation of (\ref{eq:pairingpurerewritten}) valid in the non-pure case. The final system of equations is
\begin{align}
 \label{eq:impurealmostfinal}
\fbox{$
\begin{array}{rl}
		& d_H (e^{2A-\phi} \mathrm{Re} \tilde{\psi}_1) = \pm \frac 18 e^{2A} \ast_8 \lambda (f) \ ,  \\
  & d_H (e^{2A -\phi}  \tilde{\psi}_2) =0 \ ,  \\
  & d_H^{\tilde{\mathcal{J}}_2} \bigl(e^{-\phi} \mathrm{Im} \tilde{\psi}_1 \bigr) = \pm \frac 18 f - \mathrm{Re} \bigl( \frac{2 \bar c e^\phi}{\alpha} \ast_8 \lambda (dA \wedge \tilde{\psi}_2) \bigr) + \mathrm{Re} \bigl(\gamma^{\bar{i}_1} \tilde{\psi}_2 \hat{\omega}_{\bar{i}_1} \bigr) - \mathrm{Re} \bigl(\psi_2 \gamma^{\bar{i}_2} \hat{\sigma}_{\bar{i}_2} \bigr) \ ,
\end{array}$}	
\end{align}
where the quantities $\hat{\omega}_{\bar{i}_1}$ and $\hat{\sigma}_{\bar i_2}$ are defined in (\ref{eq:omegahat}).
\section{Conclusions and future projects}
\label{sec:conclusions}

In this paper we have obtained the conditions for unbroken supersymmetry for a Mink$_2$, $\mathcal{N}= (2,0)$ vacuum in terms of generalized complex geometry. The use of the ten-dimensional system \cite{Tomasiello:2011eb} allowed us to deduce easily these conditions for a completely general vacuum, without assuming anything about the purity of the internal SUSY parameters. When specialized to the case of pure internal SUSY parameters, the system can be recast in the form (\ref{eq:susypure}) which extends the validity of \cite{Prins:2013koa} to the case of $SU(4) \times SU(4)$ structure. Our result also confirms the existence of a precise relation between the failure of the supersymmetry-calibrations correspondence and the pairing equations \cite{Martucci:2011dn}. It would be interesting to look for an extension of the supersymmetry-calibrations correspondence that take into account also the pairing equations. As a further generalization we have removed the hypothesis that $\eta^1_+$ and $\eta^2_\mp$ are pure but we have continued to assume that the structure group is $SU(4) \times SU(4)$. In this way we were able to rewrite SUSY equations in terms of pure spinors also in this case obtaining the system (\ref{eq:impurealmostfinal}). A particularly nice feature of this system (that could be analysed from the calibrations perspective) is that the exterior equations (\ref{eq:exteriorimpurefinal}) remain unchanged, whereas the pairing equations exhibit a deformation piece with respect to the pure case.

Of course the supersymmetry-calibrations correspondence is not the only reason of interest in two-dimensional vacua. First of all there is the issue of $\mathcal{N}= (1,1)$, AdS$_2$ vacua, a particular class of solutions interesting for black holes applications: four-dimensional, non rotating, extremal black holes enjoy a near-horizon geometry of the form AdS$_2 \times \Sigma_2$ with $\Sigma_2$ denoting a Riemann surface (an example is given by the well-known Reissner-Nordstr\"{o}m black holes which enjoy a near horizon geometry of the form AdS$_2 \times S_2$). Obtaining the conditions for supersymmetric AdS$_2$ vacua in Type II supergravities would be interesting in order to study the possibility of lifting such horizons to string theory. See \cite{Gran:2013kfa} and \cite{Kunduri:2013gce} for recent works on this topic.

Another possible development is the search for AdS$_3$ solutions. Usually, when obtained from the ten-dimensional system (\ref{eq:susy10}), the equations for vacua of the kind AdS$_d \times \mathcal{M}_{10-d}$ with $d$ odd are more complicated than the corresponding ones with $d$ even. Fortunately a different strategy is viable: by considering AdS$_d$ as a warped product of Mink$_{d-1}$ and $\rr$, the equations for AdS$_d$ can be easily deduced starting from the ones for Mink$_{d-1}$ (this strategy was applied, using generalized complex geometry, in \cite{Gabella:2009wu} for AdS$_5$ vacua and recently in \cite{Apruzzi:2013yva} for AdS$_7$ vacua but it has a very long story: for example it appears, in the AdS$_3$ context, in \cite{Gauntlett:2007ph} and \cite{Figueras:2007cn}). A systematic study of AdS$_3$ solutions in Type IIB supergravity started in \cite{Kim:2005ez}, which determined the conditions for the internal geometry when the only RR field non trivial is the five-form flux. After this work a large number of explicit solutions were found (see for example \cite{Gauntlett:2006ns}), and a further development occurred in \cite{Donos:2008ug} where a non trivial three-form flux was turned on. On the other hand the situation is less studied in Type IIA. Since the pure spinor approach is substantially identical for Type IIA and for Type IIB, it is conceivable that this gap can be filled starting from the conditions for unbroken supersymmetry in Mink$_2$ expressed in terms of pure spinors.

\section*{Acknowledgments}
I would like to thank F.~Apruzzi, M.~Baggioli, M.~Fazzi, C.~Imbimbo, L.~Martucci, D.~Prins and D.~Tsimpis for interesting discussions. A special thanks goes to A.~Tomasiello for having suggested this project to me and for his guidance.
My research is supported in part by INFN, by the MIUR-FIRB grant RBFR10QS5J ``String Theory and Fundamental Interactions'', and by the MIUR-PRIN contract 2009-KHZKRX.

\appendix
\section{Massaging the pairings: the pure case}
\label{sec:appendixpure}

In this section we will show the equivalence between the equations (\ref{eq:pairingpure}) and (\ref{eq:pairingpurerewritten}); we will restrict to the IIB case since the story for IIA is identical. 

First of all we need to record some further properties about the generalized Hodge diamond (\ref{eq:hodge}) and about how the deformations of the pure spinors can be arranged into the diamond. Recalling that in type IIB $\psi_1$ and $\psi_2$ are even forms on $\mathcal{M}_8$, we see that each row in the diamond has definite parity: the first row, the third and so on contain even forms, whereas the second, the fourth and so on contain odd forms. It is also straightforward to verify that $\gamma^{\bar i_1}$ (on the left) and $\gamma^{i_2}$ (on the right) act as {\it descending} operators, whereas $\gamma^{i_1}$ (on the left) and $\gamma^{\bar i_2}$ (on the right) act as {\it raising} operators: so for example by acting with  $\gamma^{\bar i_1}$ and $\gamma^{i_2}$ on $\psi_1$ it descends to the second row, whereas by acting with $\gamma^{i_1}$ and $\gamma^{\bar i_2}$ on $\bar \psi_1$ it jumps to the eighth row.

We move to discuss the deformation issues and the recipe is very simple: $\delta \psi_i$ contains only terms of the form $\gamma^{mn} \psi_i$. Concretely this means that $\delta \psi_1$ sits in the zeroth and third row of (\ref{eq:hodge}), and $\delta \psi_2$ sits in the zeroth and third column of (\ref{eq:hodge}) (and of course an identical statement is true for complex conjugates). By combining a deformation with the action of the gamma matrices we conclude that $d_H \psi_1$ sits in the second and fourth row in the diamond, whereas $d_H \psi_2$ sits in the second and fourth column.

We can now show the equivalence between (\ref{eq:pairingpure}) and (\ref{eq:pairingpurerewritten}). Our strategy is simple: we consider (\ref{eq:pairingpurerewritten}) and the first equation in (\ref{eq:exteriorpure}) and, by expanding both on each position of the diamond, we will see that they are completely equivalent to (\ref{eq:pairingpure}) plus the first equation in (\ref{eq:exteriorpure}). 

Let us start for example with the expansion in the $\psi_1 \gamma^{i_2}$ position: the first equation in (\ref{eq:exteriorpure}) rewrites
\begin{equation}
 \bigl( \bar{\psi}_1 \gamma^{\bar i_2} , 2 dA \wedge \psi_1 e^{-\phi} + d_H (e^{-\phi} \psi_1) \bigr) = - \bigl(\bar{\psi}_1 \gamma^{\bar i_2} , \frac{\alpha}{8} f \bigr) \ ,
\end{equation}
whereas (\ref{eq:pairingpurerewritten}), using the properties summarized in (\ref{eq:eigenvalues}), reads
\begin{equation}
 \bigl( \bar{\psi}_1 \gamma^{\bar i_2} ,d_H (e^{-\phi} \psi_1) \bigr) = \bigl( \bar{\psi}_1 \gamma^{\bar i_2} ,\frac{\alpha}{8} f \bigr) \ ,
\end{equation}
and, simply by subtracting the two we obtain
\begin{equation}
\bigl( \bar{\psi}_1 \gamma^{\bar i_2} , dA \wedge \psi_1 + \frac{\alpha}{8} e^\phi f \bigr) = 0 \ ,
\end{equation}
which is precisely the third equation in (\ref{eq:pairingpure}). An identical consideration shows that by considering the expansion in the $\gamma^{\bar i_1} \psi_1$ position we simply reproduce the first equation in (\ref{eq:pairingpure}). Next we consider the expansion along $\psi_2 \gamma^{\bar{i}_2}$: the first equation in (\ref{eq:exteriorpure}) gives
\begin{equation}
 \bigl(\bar{\psi}_2 \gamma^{i_2},  d_H (e^{-\phi} \psi_1) \bigr) = - \bigl(\bar{\psi}_2 \gamma^{i_2},\frac{\alpha}{8} f \bigr) \ ,
\end{equation}
whereas (\ref{eq:pairingpurerewritten}) gives
\begin{equation}
 3 \bigl(\bar{\psi}_2 \gamma^{i_2}, d_H (e^{-\phi} \psi_1) \bigr) = \bigl(\bar{\psi}_2 \gamma^{i_2},\frac{\alpha}{8} f \bigr) \ ,
\end{equation}
and the two equations imply $\bigl(\bar{\psi}_2 \gamma^{i_2},f\bigr) = 0$ which is equivalent to the fourth equation in (\ref{eq:pairingpure}). We move to the expansion along $\gamma^{\bar i_1} \psi_1 \gamma^{i_2 j _2}$:  the pairing equations (\ref{eq:pairingpure}) say nothing about this position and indeed both the first equation in (\ref{eq:exteriorpure}) and (\ref{eq:pairingpurerewritten}) say
\begin{equation}
\bigl( \gamma^{i_1} \bar{\psi}_1 \gamma^{\bar{i}_2 \bar{j}_2}, d_H (e^{-\phi} \psi_1) - \frac{\alpha}{8} f \bigr) = 0 \ ,
\end{equation}
therefore we conclude that (\ref{eq:pairingpurerewritten}) is redundant in this position.

Identical computations can be repeated for the other positions of the diamond and so we conclude that  (\ref{eq:pairingpure}) and  (\ref{eq:pairingpurerewritten}) are equivalent as we claimed.

\section{Massaging the pairings: the non-pure case}
\label{sec:appendiximpure}

In this section we will describe how the pairing equations (\ref{eq:++1}), (\ref{eq:++2}) can be massaged in the non-pure case in order to obtain the equations (\ref{eq:pairingimpuremain}). We will discuss the equations (\ref{eq:++1}) only, since the discussion for (\ref{eq:++2}) is almost identical.

To start with, we recall that for a general Mink$_2$, $\mathcal{N}= (2,0)$ vacuum configuration the pairing equations take the form (\ref{eq:++1mukai}) and (\ref{eq:++2mukai}). Let us go to consider the first equation in (\ref{eq:++1mukai}); by putting the parametrizations (\ref{eq:bilinearspureimpure}) into the equation we obtain
\begin{align}
\label{eq:pairingappone}
  & \bigl( \gamma^m [a \bar{\tilde{\psi}}_1 + b \tilde{\psi}_1 + c \bar{\tilde{\psi}}_2 + \bar c \tilde{\psi_2}], \frac 14 dA \wedge [a \tilde{\psi}_1 + b \bar{\tilde{\psi}}_1 + c \bar{\tilde{\psi}}_2 + \bar c \tilde{\psi_2}] \mp \frac{\alpha}{8} e^{\phi} \ast_8 \lambda (f) \bigr) + \nonumber \\
  + & \bigl( \gamma^m [b \bar{\tilde{\psi}}_1 + a \tilde{\psi}_1 + c \bar{\tilde{\psi}}_2 + \bar c \tilde{\psi_2}], \frac 14 dA \wedge [b \tilde{\psi}_1 + a \bar{\tilde{\psi}}_1 + c \bar{\tilde{\psi}}_2 + \bar c \tilde{\psi_2}] \mp \frac{\alpha}{8} e^{\phi} \ast_8 \lambda (f) \bigr) = 0 \ ,
\end{align} 
where the gamma matrix $\gamma^m$ has to be intended real and we have introduced the shortcuts
\begin{align}
\label{eq:shortcuts}
  & a \equiv |c_1|^2 = A_1^2 + B_1^2 + 2 A_1B_1 \sin \theta_1 \ , \qquad b \equiv |c_2|^2 = A_1^2 + B_1^2 - 2 A_1B_1 \sin \theta_1 \ , \nonumber \\
  & c \equiv \bar c_1 c_2 = A_1^2 + B_1^2 \cos (2\theta_1) + i B^2 \sin (2\theta_1) \ , \qquad d \equiv c_1 c_2 = A_1^2 - B_1^2 + 2 i A_1 B_1 \cos \theta_1 \ , \nonumber \\
  & e \equiv c_1^2 = (A_1^2 - B_1^2 \cos (2\theta_1) + 2 A_1 B_1 \sin \theta_1) + i (2 A_1 B_1 \cos \theta_1 + 2 B^2 \sin \theta_1 \cos \theta_1) \ , \nonumber \\
  & h \equiv c_2^2 = (A_1^2 - B_1^2 \cos (2\theta_1) - 2 A_1 B_1 \sin \theta_1) + i (2 A_1 B_1 \cos \theta_1 - 2 B^2 \sin \theta_1 \cos \theta_1) \ .
\end{align}
Now by taking $\gamma^m \equiv \gamma^{i_1}$ (see footnote \ref{foot:one} for the meaning of the index $i_1$), (\ref{eq:pairingappone}) simplifies to
\begin{align}
& \bigl( \gamma^{i_1} [a \bar{\tilde{\psi}}_1 +  c \bar{\tilde{\psi}}_2 ], \frac 14 dA \wedge [a \tilde{\psi}_1   + \bar c \tilde{\psi_2}] \mp \frac{\alpha}{8} e^{\phi} \ast_8 \lambda (f) \bigr) + \nonumber \\
  + & \bigl( \gamma^{i_1} [b \bar{\tilde{\psi}}_1  + c \bar{\tilde{\psi}}_2 ], \frac 14 dA \wedge [b \tilde{\psi}_1 + \bar c \tilde{\psi_2}] \mp \frac{\alpha}{8} e^{\phi} \ast_8 \lambda (f) \bigr) = 0 \ ,
\end{align}
that is
\begin{equation}
\label{eq:pairingapponefinal}
  \bigl(\gamma^{i_1} \bar{\tilde{\psi}}_1 , \frac{a^2 + b^2}{4} dA \wedge \tilde{\psi}_1 \mp \frac{\alpha}{8} (a+b) e^{\phi} \ast_8 \lambda (f)\bigr) + 2 c \bigl(\gamma^{i_1} \bar{\tilde{\psi}}_2 , \frac{\bar{c}}{4} dA \wedge   \tilde \psi_2 \mp \frac{\alpha}{8} e^\phi \ast_8 \lambda (f)\bigr) = 0 \ .
\end{equation}

Moving to the second equation in (\ref{eq:++1mukai}), we can perform the same procedure but in this case we obtain a pair of equations: the first one is obtained by taking $m= i_1$ and the second one is obtained when $m= \bar{i}_1$
\begin{align}
\label{eq:pairingapptwofinal}
  & \bar d \bigl(\gamma^{i_1} \bar{\tilde{\psi}}_1 , \frac{b}{4} dA \wedge \tilde{\psi}_1 \mp \frac{\alpha}{8}  e^{\phi} \ast_8 \lambda (f)\bigr) + \bar e \bigl(\gamma^{i_1} \bar{\tilde{\psi}}_2 , \frac{\bar{c}}{4} dA \wedge   \tilde \psi_2 \mp \frac{\alpha}{8} e^\phi \ast_8 \lambda (f)\bigr) = 0 \ , \nonumber \\
  & \bar d \bigl(\gamma^{\bar{i}_1} \tilde{\psi}_1 , \frac{a}{4} dA \wedge \bar{\tilde{\psi}}_1 \mp \frac{\alpha}{8} e^{\phi} \ast_8 \lambda (f)\bigr) + \bar h \bigl(\gamma^{\bar{i}_1} \tilde{\psi}_2 , \frac{c}{4} dA \wedge   \bar{\tilde \psi}_2 \mp \frac{\alpha}{8} e^\phi \ast_8 \lambda (f)\bigr) = 0 \ .
\end{align}
Before to proceed we note that in the pure limit we have $a=e \neq 0$ whereas $b=c=d=h=0$; therefore in this case the equations (\ref{eq:pairingapponefinal}), (\ref{eq:pairingapptwofinal}) collapse to the first two equations in (\ref{eq:pairingpure}) that are valid in the pure case. To proceed we rewrite the first equation in (\ref{eq:pairingapptwofinal}) as
\begin{equation}
\label{eq:pairingappsupport}
   \bigl(\gamma^{i_1} \bar{\tilde{\psi}}_2 , \frac{\bar{c}}{4} dA \wedge   \tilde \psi_2 \mp \frac{\alpha}{8} e^\phi \ast_8 \lambda (f)\bigr) = - \frac{\bar d}{\bar e} \bigl(\gamma^{i_1} \bar{\tilde{\psi}}_1 , \frac{b}{4} dA \wedge \tilde{\psi}_1 \mp \frac{\alpha}{8}  e^{\phi} \ast_8 \lambda (f)\bigr) \ ,
\end{equation}
that we can put in (\ref{eq:pairingapponefinal}) and in the complex conjugate of the second equation in (\ref{eq:pairingapptwofinal}) obtaining the algebraic system
\begin{align}
\label{eq:algebraicapp}
  & \bigl(a^2 + b^2 - \frac{2 c \bar d b}{\bar e} \bigr) x + \bigl(a + b - \frac{2 c \bar d}{\bar e} \bigr) y = 0 \ , \nonumber \\
  & \bigl(  d a - \frac{h \bar d b}{\bar e} \bigr) x + \bigl(  d - \frac{ h \bar d}{\bar e} \bigr) y = 0 \ , \nonumber \\
  &x \equiv \bigl( \gamma^{i_1} \bar{\tilde{\psi}}_1 , \frac 14 dA \wedge \tilde{\psi}_1 \bigr) \ , \qquad y \equiv \bigl(\gamma^{i_1} \bar{\tilde{\psi}}_1, \mp \frac{\alpha}{8} e^\phi \ast_8 \lambda (f) \bigr) \ .
\end{align}

It can be verified that the determinant of this algebraic system vanishes and so we remain with the single equation
\begin{equation}
 x = - \frac{\bar e (a+b) - 2 c \bar d}{\bar e( a^2+ b^2) - 2 c \bar d b } \, y = - \frac {1}{ 2 \alpha} y \ ,
\end{equation}
where the last equivalence can be verified using the explicit expressions (\ref{eq:shortcuts}). Summarizing the pairing equations (\ref{eq:++1mukai}) rewrite as
\begin{align}
  & \bigl(\gamma^{i_1} \bar{\tilde{\psi}}_1, dA \wedge \tilde{\psi}_1 \mp \frac 14 e^\phi \ast_8 \lambda (f) \bigr) = 0 \ , \nonumber \\
  & \bigl(\gamma^{i_1} \bar{\tilde{\psi}}_2, \mp \frac{\alpha}{8} e^\phi \ast_8 \lambda (f) \bigr) = - \bar{c} \bigl(\gamma^{i_1} \bar{\tilde{\psi}}_2, \frac{dA}{4} \wedge \tilde{\psi_2} \bigr) - \frac{\bar d}{\bar e} \bigl(\gamma^{i_1} \bar{\tilde{\psi}}_1, \frac b4 dA \wedge \tilde{\psi_1} \mp \frac{\alpha}{8} e^\phi \ast_8 \lambda (f) \bigr) \ .
\end{align}
The same strategy can be applied of course for the equations (\ref{eq:++2mukai}).

\providecommand{\href}[2]{#2}
\end{document}